\documentclass[10pt, conference]{sig-alternate-10pt}

\usepackage{multirow}
\usepackage{amssymb,amsmath}
\usepackage{graphicx}
\usepackage{subfig}
\usepackage{algorithm}
\usepackage{algorithmic}
\usepackage{color, soul}
\usepackage{cite}
\usepackage{array}
\usepackage{cases}
\usepackage{url}
\usepackage{booktabs}

\usepackage{colortbl}
\definecolor{Gray}{gray}{0.9}
\newcounter{MANumberOfComments}
\stepcounter{MANumberOfComments}

\newcounter{LINumberOfComments}
\stepcounter{LINumberOfComments}

\begin{document}

\title{The Functional Design of Scientific Computing Library}

\author{Liang Wang\\ \\
University of Cambridge, UK}

% make the title area
\maketitle

\begin{abstract} 
Owl is an emerging library developed in the OCaml language for scientific and engineering computing. It focuses on providing a comprehensive set of high-level numerical functions so that developers can quickly build up any data analytical applications. After over one-year intensive development and continuous optimisation, Owl has evolved into a powerful software system with competitive performance to mainstream numerical libraries. Meanwhile, Owl's overall architecture remains simple and elegant, its small code base can be easily managed by a small group of developers. 

In this paper, we first present Owl's design, core components, and its key functionality. We show that Owl benefits greatly from OCaml's module system which not only allows us to write concise generic code with superior performance, but also leads to a very unique design to enable parallel and distributed computing. OCaml's static type checking significantly reduces potential bugs and accelerates the development cycle. We also share the knowledge and lessons learnt from building up a full-fledge system for scientific computing with the functional programming community.
\end{abstract}

\section{Introduction}
\label{sec:intro}

Thanks to the recent advances in machine learning and deep neural networks, there is a huge demand on various numerical tools and libraries in order to facilitate both academic researchers and industrial developers to fast prototype and test their new ideas, then develop and deploy analytical applications at large scale. Take deep neural network as an example, Google invests heavily in TensorFlow\cite{199317} while Facebook promotes their PyTorch\cite{paszke2017automatic}. Beyond these libraries focusing on one specific numerical task, the interest on general purpose tools like Python and Julia\cite{julia} also grows fast.

Python has been one popular choice among developers for fast prototyping analytical applications, one important reason is because Scipy\cite{jones2014scipy} and Numpy\cite{numpy} two libraries, tightly integrated with other advanced functionality such as plotting, offer a powerful environment which lets developers write very concise code to finish complicated numerical tasks. As a result, even for the frameworks which were not originally developed in Python (such as Caffe\cite{jia2014caffe} and TensorFlow\cite{199317}), they often provide Python bindings to take advantage of the existing numerical infrastructure in Numpy and Scipy.

On the other hand, the supporting of basic scientific computing in OCaml is rather fragmented. There have been some initial efforts (e.g., Lacaml\cite{lacaml}, Oml\cite{oml}, Pareto, and etc.), but their APIs are either too low-level to offer satisfying productivity, or the designs overly focus on a specific problem domain. Moreover, inconsistent data representation and careless use of abstract types make it difficult to pass data across different libraries. Consequently, developers often have to write a significant amount of boilerplate code just to finish rather trivial numerical tasks. As we can see, there is a severe lack of a general purpose numerical library in OCaml ecosystem. We believe OCaml per se is a good candidate for developing such a general purpose numerical library for two important reasons: 1) we can write functional code as concise as that in Python with type-safety; 2) OCaml code often has much superior performance comparing to dynamic languages such as Python and Julia. 

However, designing and developing a full-fledged numerical library is a non-trivial task, despite that OCaml has been widely used in system programming such as MirageOS\cite{Madhavapeddy:2013:ULO:2499368.2451167}. The key difference between the two is obvious and interesting: system libraries provide a lean set of APIs to abstract complex and heterogeneous physical hardware, whilst numerical library offer a fat set of functions over a small set of abstract number types.

When Owl project started in 2016, we were immediately confronted by a series of fundamental questions like: "what should be the basic data types", "what should be the core data structures", "what modules should be designed", and etc. In the following development and performance optimisation, we also tackled many research and engineering challenges on a wide range of different topics such as software engineering, language design, system and network programming, and etc.

In this paper, we will introduce Owl\footnote{https://github.com/ryanrhymes/owl}, a new general purpose numerical library for scientific computing developed in OCaml.\footnote{Owl can be installed with OPAM "\texttt{opam install owl}".} We show that Owl benefits greatly from OCaml's module system which not only allows us to write concise generic code with superior performance, but also leads to a very unique design to enable parallel and distributed computing. OCaml's static type checking significantly reduces potential bugs and accelerate the development cycle. We would like to share the knowledge and lessons learnt from building up a full-fledge system for scientific computing with the functional programming community.

\section{Architecture}
\label{sec:arch}

Owl is a complex library consisting of numerous functions (over 6500 by the end of 2017), we have strived for a modular design to make sure that the system is flexible enough to be extended in future. In the following, we will present its architecture briefly.

\subsection{Hierarchy}
\label{sec:arch:hierarchy}

\begin{figure}[!htp]
  \centering 
\includegraphics[width=8.5cm]{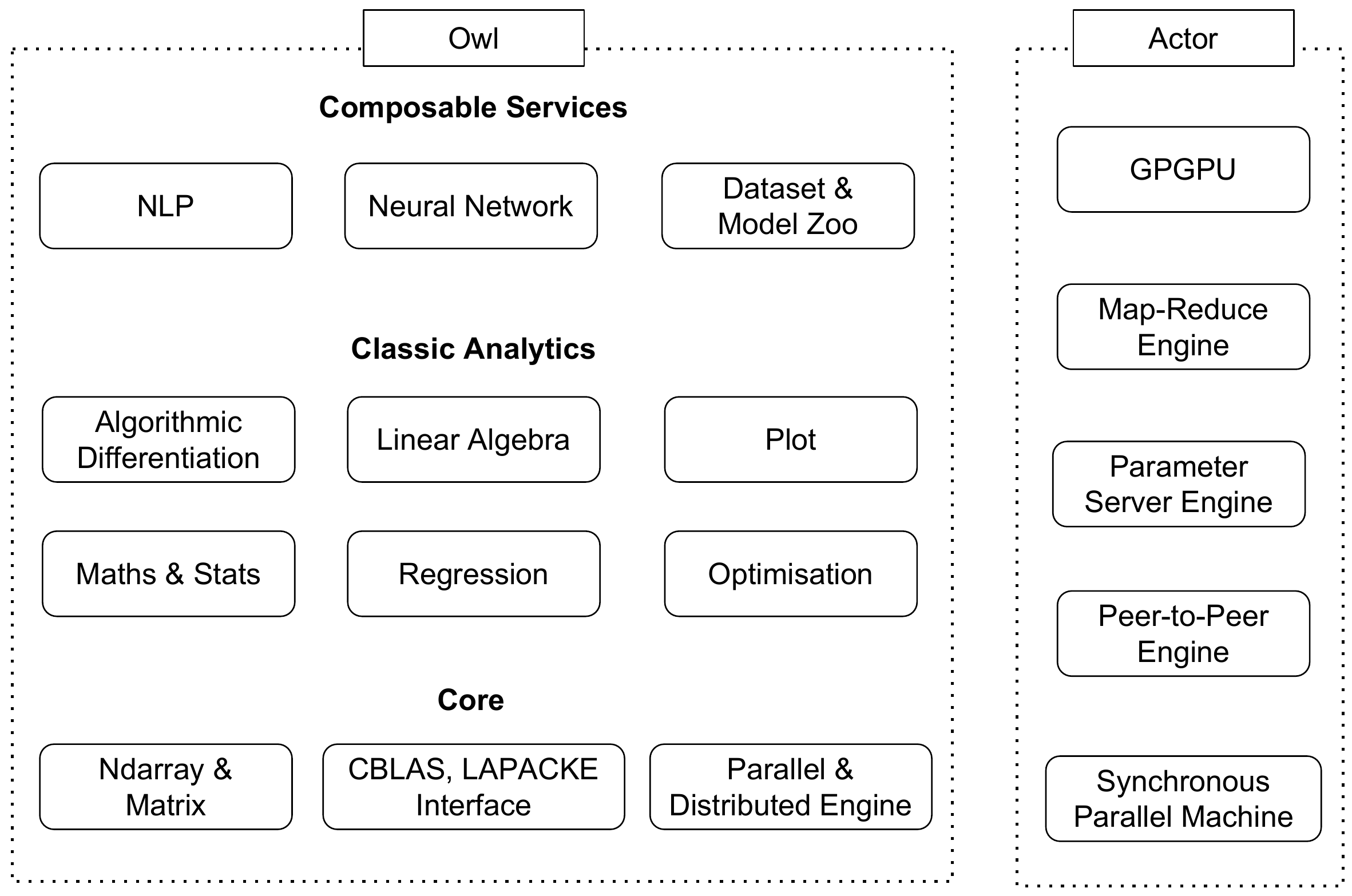}
  \caption{Two subsystems: Owl on the left handles numerical calculations while Actor on the right takes care of parallel and distributed computing.}
  \label{fig:owl_arch}
\end{figure}

Figure \ref{fig:owl_arch} gives a bird view of Owl's system architecture, i.e. the two subsystems and their modules. The subsystem on the left part is Owl's \textbf{Numerical Subsystem}. The modules contained in this subsystem fall into three categories: (1) core modules contains basic data structures and foreign function interfaces to other libraries (e.g., CBLAS and LAPACKE); (2) classic analytics contains basic mathematical and statistical functions, linear algebra, regression, optimisation, plotting, and etc.; (3) composable service includes more advanced numerical techniques such as deep neural network, natural language processing, data processing and service deployment tools.

The numerical subsystem is further organised in a stack of smaller libraries, as follows.

\begin{itemize}

\item \textbf{Base} is the basis of all other libraries in Owl. Base defines core data structures, exceptions, and part of numerical functions. Because it contains pure OCaml code so the applications built atop of Base can be safely compiled into native code, bytecode, Javascript, even into unikernels. Fortunately, majority of Owl's advanced functions are implemented in pure OCaml.

\item \textbf{Owl} is the backbone of the numerical subsystem. It depends on Base but replaces some pure OCaml functions with C implementations (e.g. vectorised math functions inNdarray module). Mixing C code into the library limits the choice of backends (e.g. browsers and MirageOS) but gives us significant performance improvement when running applications on CPU.

\item \textbf{Zoo}\cite{zhao2018privacy} is designed for packaging and sharing code snippets among users. This module targets small scripts and light numerical functions which may not be suitable for publishing on the heavier OPAM system. The code is distributed via gists on Github, and Zoo is able to resolve the dependency and automatically pull in and cache the code. 
%%% We will elaborate the design with usecases in Section \ref{sec:zoo}.

\item \textbf{Top} is the Toplevel system of Owl. It automatically loads both Owl and Zoo, and installs multiple pretty printers for various data types.

\end{itemize}

The subsystem on the right is called \textbf{Actor Subsystem}\cite{wang2017probabilistic} which extends Owl's capability to parallel and distributed computing. The addition of Actor subsystem makes Owl fundamentally different from mainstream numerical libraries such as Scipy and Julia. The core idea is to transform a user application from sequential execution mode into parallel mode (using various computation engines) with minimal efforts. The method we used is to compose two subsystems together with functors to generate the parallel version of the module defined in the numerical subsystem. We will elaborate this idea and its design in detail in Section \ref{sec:actor}.

%%%\subsection{Functor stack}
%%%\label{sec:arch:functor}
%%%
%%%OCaml's powerful module system has helped us in significantly reducing the code base of Owl. Majority of 

\section{Core data structure}
\label{sec:ndarray}

N-dimensional array and matrix are the building blocks of Owl library, their functionality are implemented in \texttt{Ndarray} and \texttt{Matrix} modules respectively. Matrix is a special case of n-dimensional array, and in fact many functions in \texttt{Matrix} module call the corresponding functions in \texttt{Ndarray} directly.

For n-dimensional array and matrix, Owl supports both dense and sparse data structures. The dense data structure is built atop of OCaml's native \texttt{Bigarray} module hence it can be easily interfaced with other libraries like BLAS and LAPACK. Owl also supports both single and double precisions for both real and complex number. Therefore, Owl essentially has covered all the necessary number types in most common scientific computations.
The functions in \texttt{Ndarray} modules can be divided into following three groups. 

\begin{itemize}

\item The first group contains the vectorised mathematical functions such as \texttt{sin}, \texttt{cos}, \texttt{relu}, and etc.

\item The second group contains the high-level functionality to manipulate arrays and matrices, e.g., \texttt{index}, \texttt{slice}, \texttt{tile}, \texttt{repeat}, \texttt{pad}, and etc. 

\item The third group contains the linear algebra functions specifically for matrices. Almost all the linear algebra functions in Owl call directly the corresponding functions in CBLAS and LAPACKE.

\end{itemize}

Aforementioned three groups of functions together provide a strong support for developing high-level numerical functions. Especially the first two groups turn out to be extremely useful for writing machine learning and deep neural network applications. Function polymorphism is achieved using GADT (Generalized algebraic data type), therefore most functions in \texttt{Generic} module accept the input of four basic number types.

\subsection{Primitives}
\label{sec:primitive}

Ndarray module has three primitives, i.e. \texttt{map}, \texttt{fold}, and \texttt{scan}. The latter two accept the \texttt{axis} parameter to specify the dimentionality along which the operation shall be performed. Practically all the vectorised math functions can be implemented using these three primitives, as follows.

\begin{itemize}

\item \texttt{map}: \texttt{ceil}, \texttt{sin}, \texttt{cos}, \texttt{relu} ...

\item \texttt{fold}: \texttt{sum}, \texttt{min}, \texttt{mean}, \texttt{std} ...

\item \texttt{scan}: \texttt{cumsum}, \texttt{cumprod}, \texttt{cummin} ...

\end{itemize}

Besides vectorised math, the three primitives also accept user-defined functions so they are sufficient to implement many other more complicated higher-order functions.

\subsection{Indexing and slicing}
\label{sec:slice}

Indexing and slicing are arguably the most important and fundamental functions in all numerical libraries. For basic slicing, each dimension in the slice definition must be defined in the format of \texttt{[start:stop:step]}. Owl provides two functions \texttt{get\_slice} and \texttt{set\_slice} to retrieve and assign slice values respectively. If the slice definition cannot be expressed by the basic format as \texttt{[start:stop:step]}, Owl provides another two functions  \texttt{get\_fancy} and \texttt{set\_fancy} that accept more complicated definition.

Because Owl's Ndarray is based on Genarray, the indexing grammar like \texttt{x.\{i, j, k ...\}} only applies to the ndarrays with dimensionality greater than four. Owl exploits the extended indexing operators recently introduced in OCaml 4.06 to provide a unified indexing and slicing interface.

\begin{itemize}

\item \texttt{.\%\{ \}} \texttt{get}

\item \texttt{.\%\{ \}<-} \texttt{set}

\item \texttt{.\$\{ \}} \texttt{get\_slice}

\item \texttt{.\$\{ \}<-} \texttt{set\_slice}

\item \texttt{.\!\{ \}} \texttt{get\_fancy}

\item \texttt{.\!\{ \}<-} \texttt{set\_fancy}

\end{itemize}

For example, the code snippet \ref{alg:05} demonstrates how to use these operator to manipulate slices in an ndarray. The grammar for slicing unfortunately still seems heavier than that of the Bigarray's native indexing. However, with the new proposal of the extended indexing introduced in OCaml 4.08\footnote{https://github.com/ocaml/ocaml/pull/1726}, the aforementioned slicing operator becomes much lighter as \texttt{x.\$\{[0;4]; [6;-1]; [-1;0]\}}, similarly for indexing grammar.

\begin{algorithm}[!tb]
  \caption{Get and Set a slice in an ndarray}
  \label{alg:05}
  \begin{algorithmic}[1]
  	\STATE{let x = sequential [|10; 10; 10|];;}
    \STATE{let a = x.\$\{ [[0;4]; [6;-1]; [-1;0]] \};;}
    \STATE{x.\$\{ [[0;4]; [6;-1]; [-1;0]] \} <- zeros (shape a);;}

  \end{algorithmic}
\end{algorithm}

\subsection{Broadcasting operation}
\label{sec:broadcast}

In numerical applications, we often need to operate ndarrays of different shapes together, e.g. adding a vector to a matrix. Conceptually, this is achieved by tiling the vector into a full matrix so that two operands have the same shape. In practice, tiling consumes more memory and can cause non-negligible burdens on GC. Broadcasting represents the trade-off between memory footprint and execution speed.

Broadcasting does not require tiling and is implicitly supported by majority of the binary math operators in Owl. If the two operands have different number of dimensions, the one of less dimensions will first be expanded using efficient \texttt{reshape} function. Then the broadcasting operation checks whether the each dimension of two ndarrays complies with the following rules: 1) both are equal; 2) either is one.

If two operands are of the same shape, the more efficient functions (with less embedded \texttt{for} loops) will be triggered to guarantee the best performance.

\subsection{Operator design}
\label{sec:operator}

OCaml does not support operator overloading at the moment which poses a big challenge when designing the operators in Owl. All the operators are centrally defined in a functor called \texttt{Owl\_operator}, which allows us to plug in different number types and avoids maintaining redundant definitions in the same code base.

Basic operators does not allow interoperation on different number types. On one hand, it is desirable to use a set of unified and consistent operators so that we can write concise code without inserting too much code merely for casting data types. On the other hand, we might not be able to take fully advantage of the static type checking since different number types are wrapped into the unified type using different type constructors.

We have been leaning more towards readability and convenience when designing Owl's operators. However, this does not necessarily mean we have to sacrifice the static type checking and performance in the library. The operators which allow interoperation on different number types are implemented in a specific module called \textit{Ext}. Each operator in \texttt{Ext} corresponds to several functions (as well as same name operators) defined in the modules of specific types.  It means we can always choose to call the functions of explicit type directly instead of using unified operators for performance and type checking concerns. 

Operators are not meant to replace those functions, which leaves us to decide how and when to use them. By the time of writing, Owl has implemented over 20 common operators such as \texttt{+}, \texttt{-}, \texttt{*}, \texttt{/}, \texttt{>}, \texttt{<}, \texttt{>.}, \texttt{\%}, and etc.

\subsection{Performance critical code}
\label{sec:cstub}

In Base library, all aforementioned basic operations such as slicing, broadcasting, and vectorised math operators have pure OCaml implementations. These OCaml functions are replaced with sequential C implementation in the core library. Since version 0.3.7, Owl has included the initial support for OpenMP with a configurable compiling switch. If the OpenMP switch is turned on in building, the sequential C code is further replaced by the parallel OpenMP implementations then compiled into the library.

\section{Computation Graph}
\label{sec:algodiff}

Computation graph plays a critical role in our system, e.g. algorithmic differentiation, lazy evaluation, and GPU computing modules all implicitly or explicitly use computation graph to perform calculations. We will present the design of these modules over this topic.

\subsection{Dynamic vs static graph}
\label{sec:compgraph}

Computation graph has a close connection to dataflow programming\cite{Johnston:2004:ADP:1013208.1013209}. There are two ways of implementing computation graphs, i.e. static graph and dynamic graph. The fundamental difference is that the structure of static graph is already determined during compilation phase whereas the dynamic graph is dynamically constructed during the runtime.

Both solutions have pros and cons. For static graph, because the structure is already known beforehand, it is easier to perform various (graph) structural optimisation and pre-allocate the memory during compilation which leads to better runtime performance. However, it is less intuitive to introduce control flow such as loop and if-else in static graphs, special nodes structure is often required. TensorFlow and Theano\cite{bastien2012theano} are typical examples of those using static graph.

For dynamic graph, the structure is constructed on the fly during the runtime. This certainly introduces runtime overhead and makes it difficult to optimise the graph structure. However, dynamic graph can be implicitly supported by operator overloading and introducing control flow is trivial since native language construct such as \texttt{while}, \texttt{for}, and \texttt{if} can be directly used in programming without any modification. PyTorch\cite{paszke2017automatic} is the typical example building atop of dynamic graph.

\subsection{Lazy Evaluation and Dataflow}
\label{sec:lazy}

Different from Haskell, OCaml eagerly evaluates the expressions. However, laziness can be very valuable in some numerical applications which involve complex computation graphs and large data chunks such as matrices. Laziness can be exploited to potentially reduce unnecessary memory allocations, optimise computation graph, incrementally update results, and etc.

Owl is able to simulate the lazy evaluation using its \texttt{Lazy} functor. Lazy functor is optimised for memory allocation in numerical computing. To reuse the allocated memory, Owl includes many inpure functions for in-place modification. E.g., \texttt{Arr.sin} is a pure function and always returns a new ndarray whilst \texttt{Arr.sin\_} performs in-place modification and overwrites the original ndarray. However, these inpure functions should not be called directly by a programmer rather than being used internally by Owl.

Lazy functor automatically tracks the reference number of each node in the computation graph and tries to reuse the allocated memory as much as possible, same as the technique called \textit{linear types} used by compilers. By so doing, Lazy module often brings us a huge benefit in both performance and memory usage. We can safely construct very complicated computation graphs without worrying about exhausting the memory. Lazy functor offers the feature of self-adjusting computing, the graph can be incrementally evaluated if only part of it needs to be updated.

\subsection{Algorithmic Differentiation}
\label{sec:algodiff}

Atop of the core components, we have developed several modules to extend Owl's numerical capability. E.g., \texttt{Maths} module includes many basic and advanced mathematical functions, whist \texttt{Stats} module provides various statistical functions such as random number generators, hypothesis tests, and so on. The most important extended module is \texttt{Algodiff}, which implements the algorithmic differentiation functionality\cite{griewank2008evaluating}. Owl’s Algodiff module is based on the core nested automatic differentiation algorithm and differentiation API of DiffSharp\cite{JMLR:v18:17-468}, which provides support for both forward and reverse differentiation and arbitrary higher-order derivatives.

The \texttt{Algodiff} module essentially bridges the gap between low-level numerical functions and high-level advanced analytical tasks such as optimisation, regression, and machine learning. Owl's algorithmic differentiation module implements both forward and backward mode (a.k.a reverse mode) and supports arbitrarily higher-order derivatives. Because \texttt{Algodiff.Generic} is implemented as a functor that takes the core data structure \texttt{Ndarry} module as inputs, it supports both single and double precisions. We will also introduce the support on complex numbers in future.

With such capability, developing advanced data analytical applications becomes trivial. E.g., we can write up the core back-propagation algorithm in neural network from scratch in just 13 lines of OCaml code. In fact we have already included a specific module in Owl for developing applications of deep neural networks. The code snippet (\ref{alg:01}) implements a simple feedforward neural network which can be used to recognise hand-writing digits. Moreover, we can also use Owl to build more complicated neural networks with graph topology such as Google's Inception network.

\begin{algorithm}[!tb]
  \caption{Build a Feedforward neural network}
  \label{alg:01}
  \begin{algorithmic}[1]
    \STATE{let network = input [| 784 |]}
    \STATE{\quad |> linear 300 $\sim$act\_typ:Activation.Tanh}
    \STATE{\quad |> linear 100 $\sim$act\_typ:Activation.Softmax}
    \STATE{in train network x y}
  \end{algorithmic}
\end{algorithm}

\begin{figure}[!htp]
  \centering 
\includegraphics[width=8.0cm]{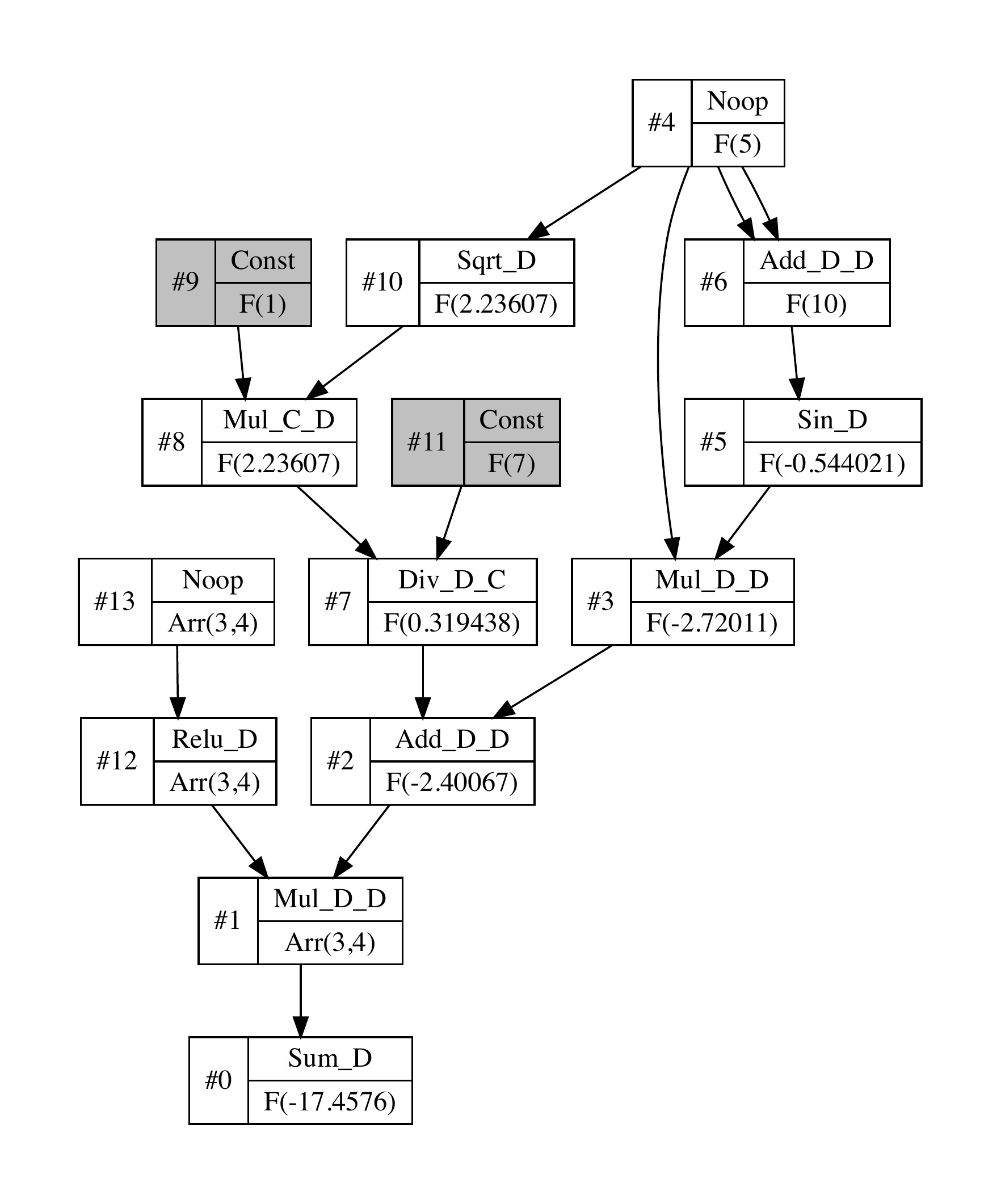}
  \caption{The computation graph generated by differentiating a simple math function in reverse mode.}
  \label{fig:compgraph01}
\end{figure}

\begin{figure}[!htp]
  \centering 
\includegraphics[width=8.0cm]{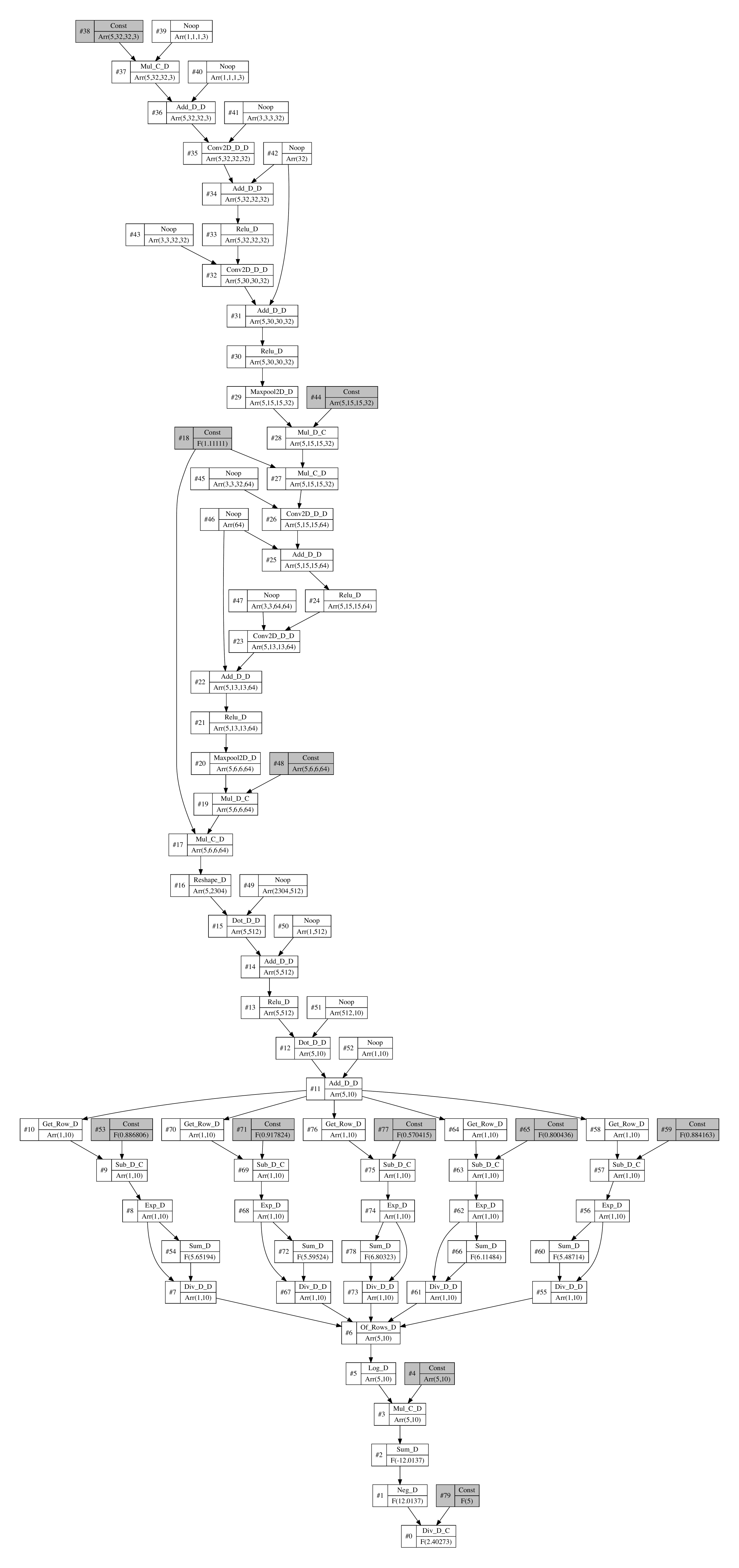}
  \caption{The computation graph generated from a VGG-like convolution neural network. Arrows represent data flows and nodes represents operations on the data. Each node contains the information on data type, shape of ndarray, type of operation, and a unique index. This information is valuable in debugging complicated numerical computations.}
  \label{fig:compgraph02}
\end{figure}

Owl is able to visualise the computation graph used in the algorithmic differentiation. The computation graph contains the detailed information of operators and operands, including the type information and shape of the data. The Fig.~\ref{fig:compgraph01} demonstrates the computation graphs from a simple math function \texttt{let f x y = Maths.((x * sin (x + x) + ( F 1. * sqrt x) / F 7.) * (relu y) |> sum)}, whist Fig~\ref{fig:compgraph02} shows the one from a VGG-like neural network.

\subsection{Optimisation engine}
\label{sec:optengine}

\texttt{Algodiff} module is able to provide the derivative, Jacobian, and Hessian of a large range of functions, we exploits this power to further build the optimisation engine. The optimisation engine is light and highly configurable, and also serves as the foundation of Regression module and Neural Network module because both are essentially mathematical optimisation problems.

The flexibility in optimisation engine leads to an extremely compact design and small code base. For a full-fledge deep neural network module, we only use about 2500 LOC and its inference performance on CPU is superior to specialised frameworks such as TenserFlow and Caffee. The code snippet \ref{alg:06} is taken from \texttt{Regression} module and implements Lasso regression. As we can see, the implementation is merely the configuration of the optimisation by specifying loss function, gradient function, regularisation term, and etc, which similarly applies to other regression models such as \texttt{ols}, \texttt{ridge}, \texttt{svm}, and etc.

\begin{algorithm}[!htb]
  \caption{Implementation of lasso regression}
  \label{alg:06}
  \begin{algorithmic}[1]
    \STATE{let lasso ?(i=false) ?(alpha=0.001) x y =}
    \STATE{\quad let params = Params.config}
    \STATE{\quad\quad $\sim$batch:Full $\sim$loss:Quadratic $\sim$gradient:GD}
    \STATE{\quad\quad $\sim$regularisation:(L1norm alpha)}
    \STATE{\quad\quad $\sim$learning\_rate:(Adagrad 1.)}
    \STATE{\quad\quad $\sim$stopping:(Const 1e-16) 1000.}
    \STATE{\quad in}
    \STATE{\quad \_linear\_reg i params x y}

  \end{algorithmic}
\end{algorithm}

\section{Parallel Computing}
\label{sec:pdc}

Parallelism can take place at various levels, e.g. on multiple cores of the same CPU, or on multiple CPUs in a network, or even on a cluster of GPUs. OCaml official release only supports single threading model at the moment, and the work on Multicore OCaml\cite{dolan2015effective} is still ongoing in the Computer Lab in Cambridge. In the following, we will present how parallelism is achieved in Owl to speed up numerical computations.

\subsection{Actor Subsystem}
\label{sec:actor}

The design of distributed and parallel computing module essentially differentiates Owl from other mainstream numerical libraries. For most libraries, the capability of distributed and parallel computing is often implemented as a third-party library, and the users have to deal with low-level message passing interfaces. However, Owl achieves such capability through its Actor subsystem.

The Actor subsystem implements three computation engines to handle both data-parallel and model-parallel computational jobs, as the following list shows.

\begin{enumerate}

\item \textbf{Map-Reduce}: implements APIs like \texttt{map}, \texttt{reduce}, \texttt{collect}, and etc. The engine is for data-parallel computational jobs.

\item \textbf{Parameter Server}: implements APIs like \texttt{register}, \texttt{push}, \texttt{pull}, and etc. The engine is for model-parallel jobs.

\item \textbf{Peer-to-Peer}: very similar to Parameter Server engine but without centralised server, each node in the system can serve part of the model.

\end{enumerate}

Owl's numerical functionality is "glued" with the Actor's distribution functionality using a well-defined function to achieve distributed analytics. The users do not need to deal with low-level message passing since this is already implemented in the engines. The users only need to focus on the high-level application logic by selecting the proper engine to use.

Actor's engine can be used to distributed and parallelise both low-level and high level data structures. E.g., the Line 1 and 2 in code\ref{alg:07}  generates a distributed Ndarray module by combining Owl's native Ndarray module and Actor's map-reduce engine.

\begin{algorithm}[!htb]
  \caption{Implementation of lasso regression}
  \label{alg:07}
  \begin{algorithmic}[1]
    \STATE{module M1 = Owl\_parallel.Make}
    \STATE{\quad (Dense.Ndarray.S) (Actor.Mapre)}
    \STATE{}
    \STATE{module M2 = Owl\_neural\_parallel.Make}
    \STATE{\quad\quad (Neural.S.Graph) (Actor.Param)}

  \end{algorithmic}
\end{algorithm}

Similarly, this method also applies to advanced data structure like neural network. Line 4 and 5 generates a new neural network module which enables distributed training by combining Owl's Neural module and Actor's Parameter Server engine.

\subsection{GPU Computing}
\label{sec:gpu}

Scientific computing involves intensive computations, and GPU has become an important option to accelerate these computations by performing parallel computation on its massive cores. There are two popular options in GPGPU computing: CUDA and OpenCL. CUDA is developed by Nvdia and specifically targets their own hardware platform whereas OpenCL is a cross platform solution and supported by multiple vendors.  Owl currently supports OpenCL and CUDA support is included in our future plan.

\begin{figure}[!htp]
  \centering 
\includegraphics[width=8.0cm]{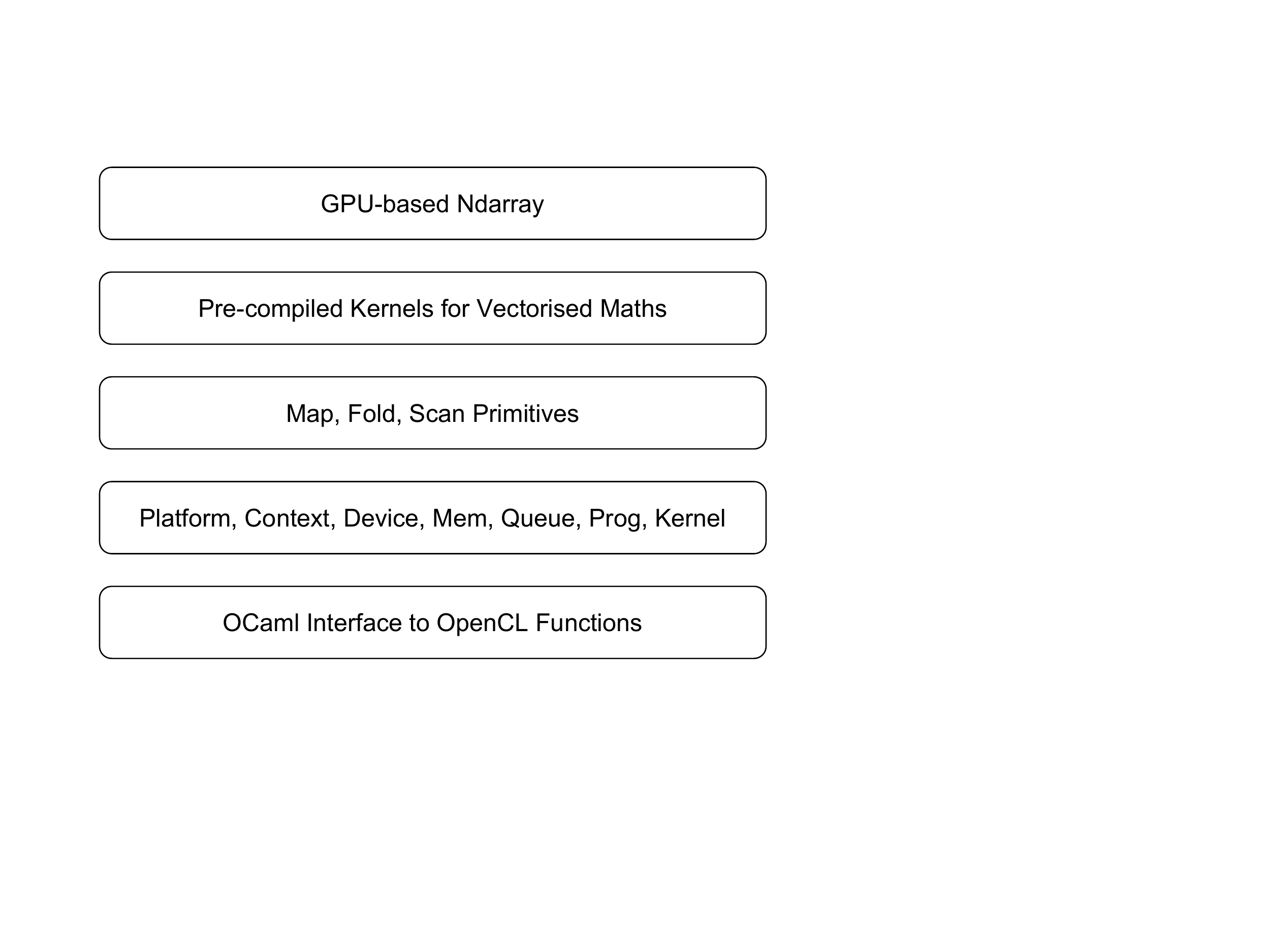}
  \caption{Module stack of Owl's OpenCL sublibrary.}
  \label{fig:opencl}
\end{figure}

The OpenCL support is implemented as a sub-library, and Figure \ref{fig:opencl} presents its module stack. The bottom two layers re-wrap OpenCL objects (e.g. Platform, Context, Device, and etc.) into modules using OCaml interface to OpenCL C functions. The middle layer implements \texttt{map}, \texttt{fold}, and \texttt{scan} three primitive with which a comprehensive set of math functions (e.g. \texttt{add}, \texttt{sub}, \texttt{mul}, \texttt{sin}, etc.) is implemented as default kernels. There kernels are pre-compiled every time when Owl library is loaded.

\begin{algorithm}[!tb]
  \caption{Perform computation on GPU and CPU}
  \label{alg:02}
  \begin{algorithmic}[1]
  	\STATE{let cpu\_compute x =}
    \STATE{\quad mul x x |> sin |> cos |> tan |> neg}
    \STATE{\quad }
    \STATE{let gpu\_compute x =}
    \STATE{\quad let open Owl\_opencl\_dense\_ndarray in}
    \STATE{\quad let x = of\_ndarray x in}
    \STATE{\quad let y = mul x x |> sin |> cos |> tan |> neg in}
    \STATE{\quad to\_ndarray float32 y}

  \end{algorithmic}
\end{algorithm}

In code snippet \ref{alg:02}, the two functions run the same calculation on CPU and GPU respectively. Similar to Lazy module, OpenCL module internally maintains a computation graph for an expression. The expression is lazily evaluation to reduce the number of memory copying between the host and GPU devices. \texttt{of\_ndarray} converts a normal CPU ndarray to a GPU ndarray, whilst \texttt{to\_ndarray} implicitly calls \texttt{eval} function and copies the result back to the host memory.

Considering a user may have implemented many functions previously in OpenCL C code, Owl also provides flexible mechanisms to allow the user to include the OpenCL code and run it directly on OCaml's native Bigarray. As presented in code snippet \ref{alg:03}, line 3 to 7 is an OCaml native string containing the OpenCL C code which adds up two arrays. These raw OpenCL code will be dynamically compiled during the runtime by calling \texttt {add\_kernels} function then the compiled kernel is included in Owl's kernel cache to avoid recompilation in future invocations.

\begin{algorithm}[!tb]
  \caption{Mixing OpenCL code with OCaml code}
  \label{alg:03}
  \begin{algorithmic}[1]
  	\STATE{let add\_two\_array () =}
    \STATE{\quad let code = "}
    \STATE{\quad \quad \_\_kernel void add\_xy}
    \STATE{\quad \quad \quad (\_\_global float *a, \_\_global float *b) \{}
    \STATE{\quad \quad \quad int gid = get\_global\_id(0);}
    \STATE{\quad \quad \quad a[gid] = a[gid] + b[gid];}
    \STATE{\quad \quad \}"}
    \STATE{\quad in}
    \STATE{\quad Context.(add\_kernels default [|code|]);}
    \STATE{\quad let x = Dense.Ndarray.S.uniform [|20;10|] in}
    \STATE{\quad let y = Dense.Ndarray.S.sequential [|20;10|] in}
    \STATE{\quad Context.eval $\sim$param:[|F32 x; F32 y|] "add\_xy"}

  \end{algorithmic}
\end{algorithm}

Line 12 invokes the OpenCL function by passing the its name to \texttt{Context.eval} and corresponding parameters. Note that the OpenCL function names must be unique since they are all maintained in a hash table.

\subsection{OpenMP}
\label{sec:openmp}

OpenMP uses shared memory multi-threading model to provide parallel computation. It is requires both compiler support and linking to specific system libraries. OpenMP support is transparent to programmers, it can be enabled by turning on the corresponding compilation switch in \textbf{jbuild} file. 

After enabling OpenMP, many vectorised math operators are replaced with the corresponding OpenMP implementation in the compiling phase. Parallelism offered by OpenMP is not a free lunch, the scheduling mechanism adds extra overhead to a computation task. If the task per se is not computation heavy or the ndarray is small, OpenMP often slows down the computation. We therefore set a threshold on ndarray size below which OpenMP code is not triggered. This simple mechanism turns out to be very effective in practice.

\section{Interfaced Libraries}
\label{sec:utility}

Some functionality such as plotting and linear algebra is included into the system by interfacing to other libraries. Rather than simply exposing the low-level functions, we carefully design easy-to-use high-level APIs and this section will cover these modules.

%%%\subsection{Testing Framework}
%%%\label{sec:testing}

%%%\subsection{Zoo Subsystem}
%%%\label{sec:zoo}

\subsection{Linear Algebra}
\label{sec:linalg}

Even though Fortran is no longer among the top choices as a programming language, there is still a large body of Fortran numerical libraries whose performance still remain competitive even by today's standard, e.g. BLAS and LAPACK.

When designing the linear algebra module, we decide to interface to CBLAS and LAPACKE (i.e. the corresponding C interface of BLAS and LAPACK) then further build higher-level APIs atop of the low-level Fortran functions. The high-level APIs hides many tedious tasks such as setting memory layout, allocating workspace, calculating strides, and etc.

One thing worth noting is that Owl only uses C layout (i.e. row-based layout) to represent ndarray in the memory. This decision is based on two considerations: first, using one layout significantly reduces the code base, we do not have to implement two separate versions of functions for each ndarray operation in order to obtain the best performance. Second, using Fortran layout indicates indexing starts from 1 whereas OCaml's native indexing (e.g. list and array) starts from 0. Mixing two indexing styles turns out to be a source of bugs in our experience.

\subsection{Plotting Functions}
\label{sec:plot}

Plotting is an indispensable function in modern numerical libraries. We build Plot module on top of PLplot\cite{lebrun2005plplot} which is a powerful cross-platform plotting library. However PLPlot only provides very low-level functions to interact with its multiple plotting engines, eveb making a simple plot involves very lengthy and tedious control sequence. Using these low-level functions directly requires developers to understand the mechanisms in depth, which not only significantly reduces the productivity but also is prone to errors.

Inspired by Matlab, we implement Plot module to provide developers a set of high-level APIs. The core plotting engine is very lightweight and only contains about 200 LOC. Its core design is to cache all the plotting operations as a sequence of function closures and execute them all when we output the figure.

\begin{figure}[!htp]
  \centering 
\includegraphics[width=8.5cm]{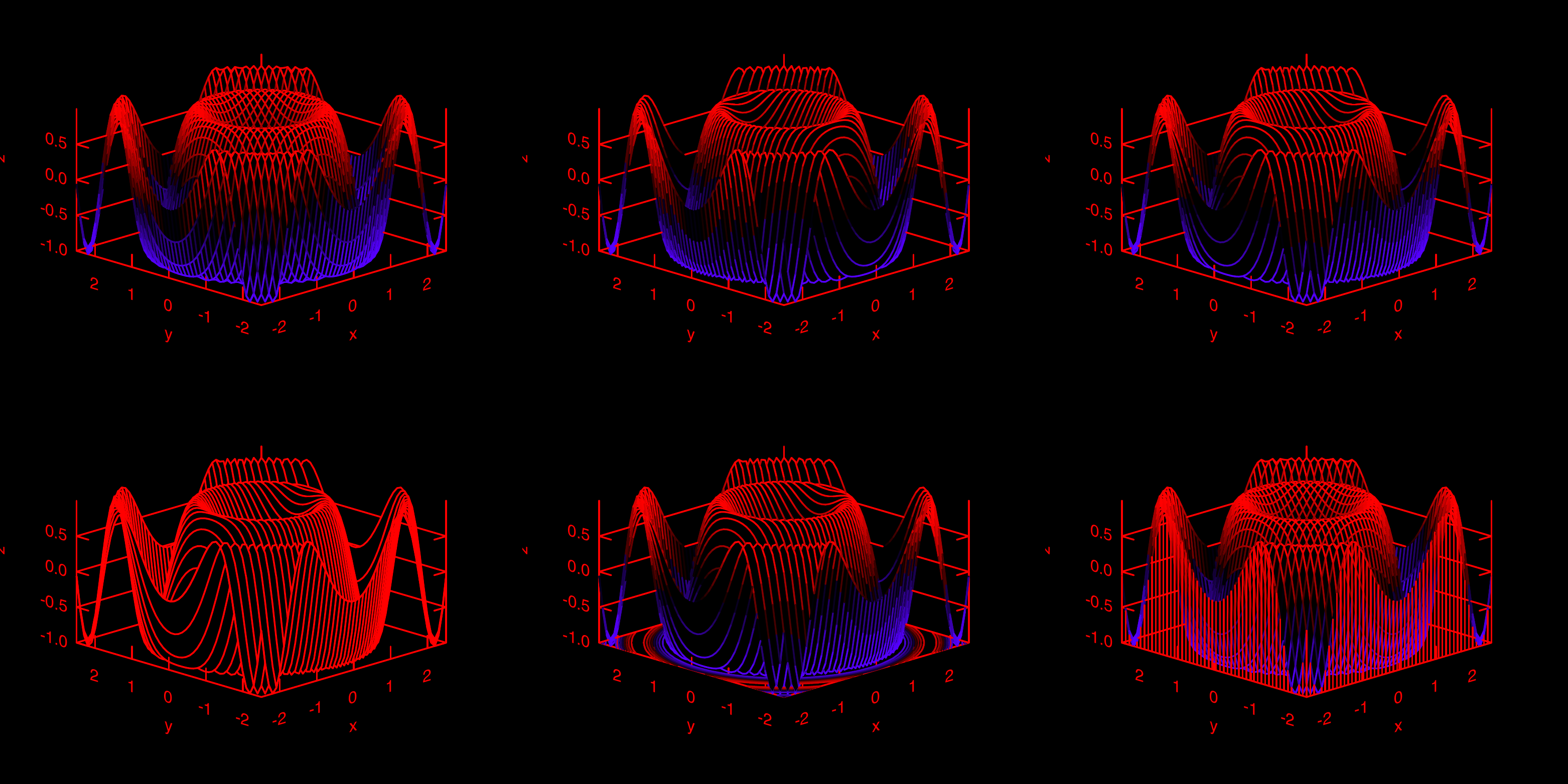}
  \caption{Subplots generated by Owl's Plot module.}
  \label{fig:subplot}
\end{figure}

The Figure \ref{fig:subplot} presents a subplot generated by the code snippet \ref{alg:04}. The \texttt{Plot.create} function on the third line returns a handle of a $2 \times 3$ subplot. Then the following calls on \texttt{Plot.subplot i j} shifts the "focus" on each subplot with specified coordinates $(i,j)$ to let the programmer perform further plotting operations. The \texttt{Plot.output} function on Line 16 evaluates all the cached function closures on the canvas and output the final figure.

\begin{algorithm}[!htb]
  \caption{Using Plot module to make subplots}
  \label{alg:04}
  \begin{algorithmic}[1]
    \STATE{let make\_subplots x y z =}
    \STATE{\quad let open Plot in}
    \STATE{\quad let h = create $\sim$m:2 $\sim$n:3 "demo.pdf" in}
    \STATE{\quad subplot h 0 0;}
    \STATE{\quad mesh $\sim$h $\sim$spec:[ ZLine XY ] x y z;}
    \STATE{\quad subplot h 0 1;}
    \STATE{\quad mesh $\sim$h $\sim$spec:[ ZLine X ] x y z;}
    \STATE{\quad subplot h 0 2;}
    \STATE{\quad mesh $\sim$h $\sim$spec:[ ZLine Y ] x y z;}
    \STATE{\quad subplot h 1 0;}
    \STATE{\quad mesh $\sim$h $\sim$spec:[ ZLine Y; NoMagColor ] x y z;}
    \STATE{\quad subplot h 1 1;}
    \STATE{\quad mesh $\sim$h $\sim$spec:[ ZLine Y; Contour ] x y z;}
    \STATE{\quad subplot h 1 2;}
    \STATE{\quad mesh $\sim$h $\sim$spec:[ ZLine XY; Curtain ] x y z;}
    \STATE{\quad output h}

  \end{algorithmic}
\end{algorithm}

We introduce a named parameter called \texttt{spec} in each plotting function so that the programmer can pass in a list of specification to further configure a plot. This gives Plot module the flexibility for future extension, the unknown or unsupported specification will be ignored by the plotting function.

\section{Preliminary Evaluation}
\label{sec:eval}

%%% Performance is the key in many numerical applications, we have performed one preliminary evaluation of Owl lately, and the initial results appear promising.
Since Owl provides the underlying numerical functions for actor, we have compared its performance with several other mainstream tools. Even though TensorFlow is not a general purpose numerical library, it is included due to relatively strong support for tensor operations. The numerical library often contains thousands of APIs, therefore we cannot provides the complete comparison due to the space limit. In this paper, we only select one function from each category and present the results in table \ref{owl:perf:1}. Note all the evaluations are performed CPU only.

\begin{table*}[]
\centering
\caption{Performance comparison of core functions between different libraries. The time is measured in \textbf{ms}.}
\label{owl:perf:1}
\begin{tabular}{@{}lllll@{}}
\toprule
Operation              & Owl         & Scipy       & Julia       & TensorFlow  \\ \midrule
slicing {[}*; 0:499{]} & 0.80 (0.07) & 0.94 (0.21) & 0.78 (0.19) & 1.04 (0.04) \\
slicing {[}0:499; *{]} & 0.65 (0.04) & 0.79 (0.12) & 0.70 (0.10) & 0.84 (0.08) \\
relu (map)             & 3.51 (0.18) & 3.67 (0.24) & 3.48 (0.15) & 3.53 (0.09) \\
sum (fold)             & 0.57 (0.04) & 3.58 (0.18) & 1.84 (0.08) & 0.74 (0.06) \\
cumsum (scan)          & 6.80 (0.21) & 9.21 (0.31) & 6.53 (0.94) & 6.62 (0.84) \\
x + y                  & 3.83 (0.46) & 5.82 (0.38) & 4.39 (0.50) & 4.02 (0.10) \\
inv(x)                 & 88.5 (2.23) & 91.6 (3.38) & 95.8 (1.70) & 472  (10.1) \\
iter                   & 39.3 (1.31) & 4378 (283)  & 538  (32.2) & 13.2 (0.35) \\ \bottomrule
\end{tabular}
\end{table*}

We use Owl (version 0.3.0) in the evaluation and compare to Numpy (version 1.8.0rcl) and Julia (version 0.5.0)\cite{julia}. All evaluations are done on the same MacBook Air (1.6GHz Intel Core i5, 8GB memory).
In the evaluation, we create a $1000 \times 1000$ random matrix (uniform distribution in $[0,1]$). Each experiment is repeated 100 times and both the mean and the standard deviation (in the parenthesis) are reported. The first ten evaluations serve as warm-up phase and are excluded from calculation. This is especially important for Julia because its first run requires a significant amount of extra time for compilation.

Regarding the selected operations, we first choose slicing function since it is a fundamental operation of n-dimensional array which serves as a core data structure in all the numerical libraries. Second, many operations on n-dimensional array can be categorised as \texttt{map}, \texttt{fold} (or \texttt{reduce}), and \texttt{scan} three groups, hence we choose one function from each. Third, we include one binary math operator \texttt{+} for addition as well as \texttt{inv} for matrix inversion as basic linear algebra examples. In the last, we also include \texttt{iter} function for iterating all the elements which can be used to build many useful operations.

For both slicing and vectorised mathematical operations, we can see all the libraries achieve similar performance and in most cases Owl is the best whereas Scipy is the worst. Owl and Scipy have slightly higher standard deviation which is the effects of GC. TensorFlow offloads many functions to Eigen\footnote{http://eigen.tuxfamily.org/}, and it is interesting to notice that a library developed in C++ does not necessarily indicate better performance.

For \texttt{inv}, TensorFlow is significantly slower since it uses its native implementation whereas the rest three interfaces to Lapack(e). For iteration function, TensorFlow is much faster since the loop overhead is significantly smaller in C++ whereas the rest three are slower. However, Owl still possesses superior performance to Julia and Scipy due to OCaml highly efficient language runtime.

\begin{figure}
  \centering
  \subfloat[Inference performance]{
    \label{f:owl:2:1}\includegraphics[width=0.42\textwidth]%,valign=c]
    {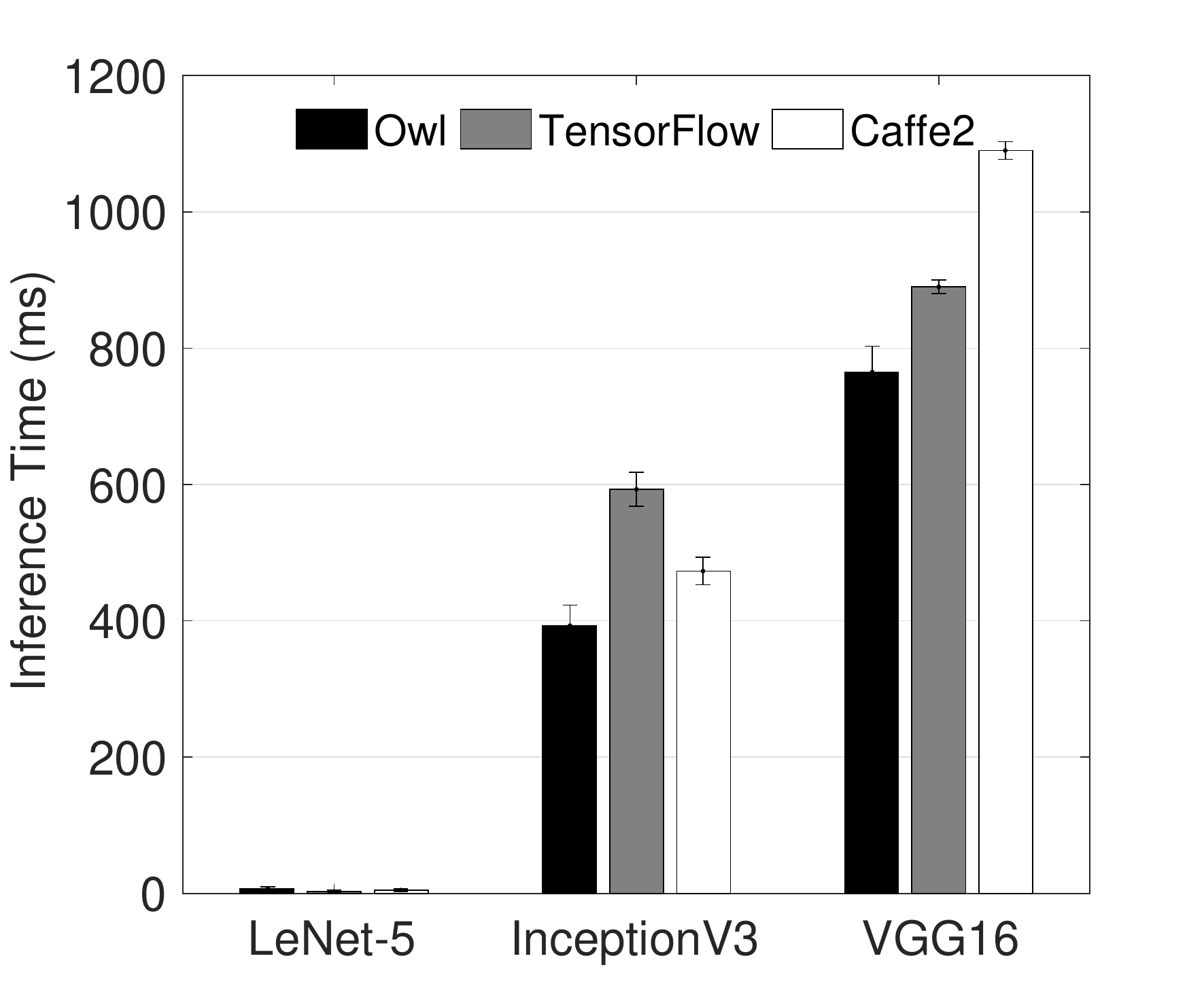}
  }
  \hfill
  \subfloat[Training performance]{
    \label{f:owl:2:2}\includegraphics[width=0.42\textwidth]%,valign=c]
    {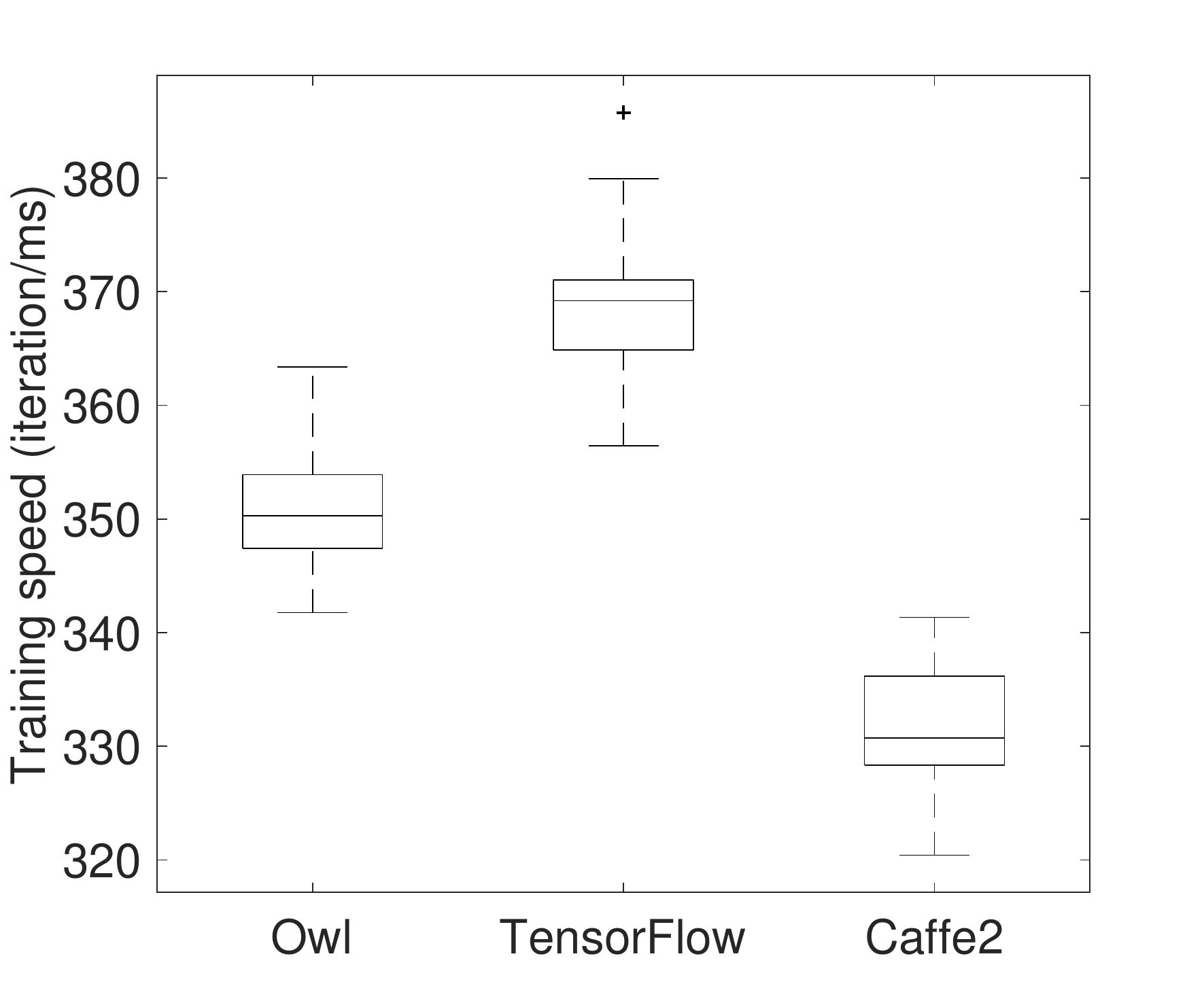}
  }
  \caption{\label{f:owl:2}Performance comparison between different deep neural network frameworks. Owl is able to achieve superior performance in both training and inference.}
\end{figure}

Besides benchmarking the basic numerical functions, we also evaluate a set of more advanced examples and report the results in Figure~\ref{f:owl:2}. In this set of evaluations, we compare Owl to two popular deep neural network frameworks Caffe2 (built on original Caffe\cite{jia2014caffe}) and TensorFlow\cite{199317}.

In Figure~\ref{f:owl:2:1}, we evaluate the inference performance of three different neural networks, i.e. LeNet-5, Google's Inception\cite{7780677}, and VGG16\cite{simonyan2014very}. Owl is able to achieve the best performance in both Inception and VGG16. All three frameworks have similar performance on LeNet, but the results are subject to higher variance due to LeNet's small size. As the neural network gets more complicated, the performance gain of using Owl becomes more noticeable.

In Figure~\ref{f:owl:2:2}, we build and train a VGG16 network using CIFAR\cite{krizhevsky2009learning} dataset. On all three frameworks, we try to keep most experiment settings the same for fair comparison, e.g. all use Adagrad 0.005 as learning rate. Due to the same setting, the convergence speed w.r.t number of iterations is the same on all frameworks, therefore we measure the time spent in each iteration as its training speed. The evaluation shows the difference between these frameworks is not significant with Caffe2 often has slightly better performance and TensorFlow has the worse one. Facebook has devoted a lot of efforts in optimising Caffe2 for production use whereas TensorFlow and Owl target remain higher flexibility as a framework for research purpose.

%%%\section{Discussion}
%%%\label{sec:discuss}
%%%
%%%Pros and cons in developing a numerical library in Functional Language, specifically in OCaml.
%%%
%%%Fast iteration in the very beginning phase of the project, this is extremely valuable. Type checking shields a lot of troubles from potential bugs. The cost of re-adjusting the system architecture becomes much lowers.
%%%
%%%Expressiveness and code reuse,  it also significantly reduces the code base and maintenance overhead.

\section{Conclusion}
\label{sec:conclusion}

In this paper, we presented Owl - an OCaml library for scientific and engineering computing. We introduced Owl's design principles, as well as its core components, key functionality, advanced algorithmic differentiation. We also provided several examples to show how to take advantage of Owl to fast prototype new ideas. Our preliminary evaluation has shown that Owl has superior performance comparing to other mainstream numerical libraries, meanwhile its architecture remains simple and the code case is small enough to be managed by a small group of developers.

% that's all folks
\end{document}